\documentclass[twoside]{article}
\usepackage{Aat}
\usepackage{graphicx}
\usepackage{tabularx}
\usepackage{Aattable}
\usepackage{amsmath, amsfonts, amssymb}

\newcommand{\FigBox}[2][\columnwidth]{\framebox[#1]{\rule{0pt}{#2}}}
\begin{document}
\thispagestyle{myfirst}
\setcounter{page}{1}
\mylabel{00}{4}
\mytitle{Supercomputer Simulations of Disk Galaxies}

\myauthor{Evgeny Griv, Michael Gedalin, Edward Liverts, David Eichler}
\myadress{Dept. of Physics, Ben-Gurion University, Beer-Sheva 84105,
Israel}
\myauthor{Yehoshua Kimhi}
\myadress{Inter University Computational Center,
Ramat Aviv 69978, Israel}
\mydate{(Received December  1, 2000)}

\myabstract{The time evolution of models for an isolated disk of highly
flattened galaxies of stars is investigated by direct integration
of the Newtonian equations of motion of $N=30,000$ identical stars 
over a time span of many galactic rotations.  Certain astronomical
implications of the simulations to actual disk-shaped (i.e. rapidly
rotating) galaxies are explored as well.}

\mykey{Galaxies ~~kinematics and dynamics ~~galaxies 
~~structure--instabilities}

One can learn much about the properties of stellar systems of disk
galaxies experimentally by computer simulation of many-body systems.
We analyze the evolution and stability of structures in $N$-body
models of isolated, rapidly and nonuniformly rotating, and
spatially inhomogeneous stellar disks
of galaxies by direct integration over a time span of Newtonian
equations of motion of identical particles.  Use of concurrent 
computers has enabled us to make long simulation runs using a
sufficiently large number of particles.  The essential difference
between the present and previous simulations is the comparison
between the results of $N$-body experiments and the stability 
theory as developed by Griv and Peter (1996), Griv {\it et al.} 
(1997, 1999a, 1999b, 2000) and Griv (1998). 

At the start of the $N$-body integration, our similation initilizes
the particles on a set of concentric circular rings with a circular
velocity $\bf V$ of galactic rotation in the equatorial plane; the
system is isolated in vacuum.  Then the position of each particle
was slightly perturbed by applying a pseudorandom number generator.
The Maxwellian-distributed random velocities $\bf v$ were added to
the initial circular velocities $\bf V$, and $|{\bf v}| \ll
|{\bf V}|$.  Finally, slight corrections have
been applied to the resultant velocities and coordinates of the
model stars so as to ensure the equilibrium between the centrifugal
and gravitational forces and to preserve the position of the disk
center of gravity at the origin.

\begin{figure}[t]
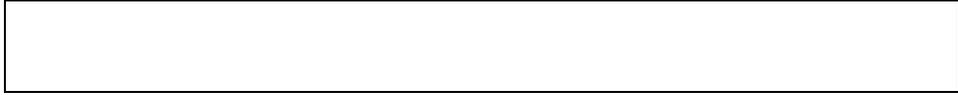

\FigBox{1cm}
\caption{The time evolution (face-on view) of 
a Jeans-unstable cold disk of $N=30,000$ stars.  
Notice how rapidly the small-scale Jeans instability grows
with time.  At the final stage the pair of strongly interacting
M51 type galaxies consisting of the main massive galaxy and a 
minor one is developed.  Note the resemblance of the structures
seen here at times $t = 0.8-2.6$ to ones which were usually 
classified (and included in the Atlas of Interacting
Galaxies) by Vorontsov-Velyaminov (1977a, 1987) as the ``nests" 
and ``chains."  Vorontsov-Velyaminov considered these objects 
as compact fragmenting systems, giving birth to young galaxies.  
Based on the present simulations, a new mechanism for the formation 
of definitely interacting galaxies in close pairs and groups may be
suggested through the fragmentation of a Jeans-unstable 
(Toomre-unstable) disk of an original giant galaxy.}
\end{figure}

\begin{figure}[t]
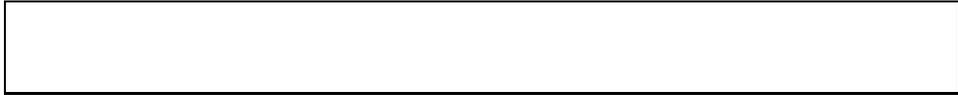

\FigBox{1cm}
\caption{Higher resolution plots (edge-on view) for the simulation
run shown in Figure 1.  The striking resemblance of the chain
structures seen in the $N$-body model at times $t=1.4-2.6$ and ones
revealed by Vorontsov-Velyaminov (1977a, 1987) (see also Zasov  
{\it et al.}, 2000) probably confirms our suggestion that these 
structures in strongly interacting close galaxies is indeed 
produced by the Jeans instability that develops in the dynamically
cold Toomre-unstable disk of stars.  Earlier, Vorontsov-Velyaminov 
(1974, 1977b, 1987) already presented observational evidence
of the fragmentation of galaxies at the present epoch.  The latter 
represents a factor of quantitative importance in the general 
evolution of galaxies themselves and of the intergalactic medium
(e.g. the density in intergalactic space becomes higher).}
\end{figure}

In Figure 1 we show a series of face-on view snapshots from a
three-dimensional simulation run of the so-called cold disk,
in which the initial dispersion of random velocities of stars
was chosen to be less than the critical Toomre's (1964) dispersion.
The time was normalized so that the time $t=1.0$ corresponds to a
single revolution of the initial disk; the rotation was taken to be
counterclockwise.  It is seen that the system is violently unstable
to small-scale gravity perturbations of the Jeans type.  During the
first rotation, such unstable perturbations break the system into
several macroscopic fragments of stars.  At the end of the second
rotation, we see a quasi-stationary binary system of strongly 
interacting galaxies.  According to Vorontsov-Velyaminov (1987),
the average estimate of the number of such strongly interacting 
pairs is $14\%$ of single galaxies.

In Figure 2 we show the time evolution of the cold disk in the
direction normal to the plane.  From an initial very thin model, a
fully three-dimensional disk develops immediately at $t \approx 0.2$
with a mean height above the plane, corresponding to the force
balance between the gravitational attraction in the plane and the
``pressure" due to the velocity dispersion (i.e. ``temperature")
in the vertical direction.  At a time $t \approx 1.4$, in Figure 
2, one can see a small chain system consisting of interacting 
galaxies.  At the same time a bending firehose type 
instability develops in each fragment.  See Griv and Chiueh 
(1998) for a discussion of the bending instability.  This
instability essentially increases the disk thickness of 
compact fragments.

\subsection*{Acknowledgements}

This work was performed in part under the auspices of the Israel
Science Foundation, the Israeli Ministry of Immigrant Absorption
and the Israel--U.S. Binational Science Foundation.  The authors
are grateful to Arthur Chernin, Tzi-Hong Chiueh, Alexei Fridman, 
Shlomi Pistinner, Raphael Steinitz and Chi Yuan for valuable
discussions.

\subsection*{References}

\rf{Griv, E., 1998, {\it Astro. Lett. Commu.}, {\bf 35}, 403.}

\rf{Griv, E. and Peter, W., 1996, {\it Astrophys. J.}, {\bf 469}, 84.}

\rf{Griv, E., Gedalin, M. and Yuan, C., 1997, {\it Astron.
   Astrophys.}, {\bf 328}, 531.}

\rf{Griv, E. and Chiueh, T., 1998, {\it Astrophys. J.}, {\bf 503}, 186.}

\rf{Griv, E., Rosenstein, B., Gedalin, M. and Eichler, D., 1999a,
   {\it Astron. Astrophys.}, {\bf 347}, 821.}

\rf{Griv, E., Yuan, C. and Gedalin, M., 1999b, {\it Month. Not. R. Astron. 
   Soc.}, {\bf 307}, 1.}

\rf{Griv, E., Gedalin, M., Eichler, D. and Yuan, C., 2000, {\it Phys. Rev.
   Lett.}, {\bf 84}, 4280.}

\rf{Toomre, A., 1964, {\it Astrophys. J.}, {\bf 139}, 1217.}

\rf{Vorontsov-Velyaminov, B. A., 1974, {\it Astron. Astrophys.}, {\bf 37},
   425.}

\rf{Vorontsov-Velyaminov, B. A., 1977a, {\it Astron. Astrophys. Suppl.
   Ser.}, {\bf 28}, 1.}

\rf{Vorontsov-Velyaminov, B. A., 1977b, {\it Soviet Astron. Lett.}, 
   {\bf 3}, 132.}

\rf{Vorontsov-Velyaminov, B. A., 1987, Extragalactic Astronomy, Harwood,
   London.}

\rf{Zasov, A. V. {\it et al.}, 2000, {\it Astron. Astrophys. Suppl. Ser.}, 
   {\bf 144}, 429.}

\end{document}